# Not One to Rule Them All: Mining Meaningful Code Review Orders From GitHub

Abir Bouraffa
University of Hamburg
Germany
abir.bouraffa@uni-hamburg.de

Carolin Brandt
Delft University of Technology
The Netherlands
c.e.brandt@tudelft.nl

Andy Zaidman
Delft University of Technology
The Netherlands
a.e.zaidman@tudelft.nl

Walid Maalej
University of Hamburg
Germany
walid.maalej@uni-hamburg.de

## Abstract
Developers use tools such as GitHub pull requests to review code, discuss proposed changes, and request modifications. While changed files are commonly presented in alphabetical order, this does not necessarily coincide with the reviewer's preferred navigation sequence. This study investigates the different navigation orders developers follow while commenting on changes submitted in pull requests. We mined code review comments from 23,241 pull requests in 100 popular Java and Python repositories on GitHub to analyze the order in which the reviewers commented on the submitted changes. Our analysis shows that for 44.6% of pull requests, the reviewers comment in a non-alphabetical order. Among these pull requests, we identified traces of alternative meaningful orders: 20.6% (2,134) followed a largest-diff first order, 17.6% (1,827) were commented in the order of the files' similarity to the pull request's title and description, and 29% (1,188) of pull requests containing changes to both production and test files adhered to a test-first order. We also observed that the proportion of reviewed files to total submitted files was significantly higher in non-alphabetically ordered reviews, which also received slightly fewer approvals from reviewers, on average. Our findings highlight the need for additional support during code reviews, particularly for larger pull requests, where reviewers are more likely to adopt complex strategies rather than following a single predefined order.

## 1 Introduction

Code review, the practice of inspecting another developer's changes before integrating them into a code base, has become a cornerstone of quality assurance in software engineering [4]. To facilitate code reviews, developers use tools like GitHub pull requests (PRs)[1], gerrit[2], or phabricator[3]. These tools present the contributed changes in a diff view for each file, where added and deleted lines are shown along with a context of unchanged lines. Usually, changed files are simply shown in the alphabetical order of their project path, i.e., the default case on GitHub. However, this default arrangement might not align with the order in which reviewers prefer to inspect code.

In their call for a new generation of code review tools, Baum and Schneider [5] advocate the need for more support in change understanding stating that reviewers attempt to find an order that helps them comprehend the change, only to fall back on the order presented in the review tool. Subsequent studies have also shown that the amount of file comments decreases for files shown later in the code review [3, 14]. Consequently, a renewed interest in software engineering research has emerged, focusing on the optimal order for displaying files during code review [3, 6, 16].

To better understand what a meaningful code review order is, this study investigates the order in which reviewers comment on changes in PRs. Our work is based on two central assumptions: 1) There are developers that review code in a different order than the default alphabetical order, despite the lack of tool support for this. 2) There is no single optimal order for code review. Instead, developers adopt different review orders depending on the context, such as the specific project or situation.

In this study, we mine code review comments from 100 popular Java and Python repositories on GitHub, and analyze how often developers' comments follow the alphabetical order. For code reviews not following that order, we investigate whether other meaningful orders can explain the deviation. Particularly, we check test first, largest-diff first, and most-similar first orders, the latter referring to how similar the file diff is with the pull request title and description. We derive meaningful orders from existing literature proposing an

---
[1] https://github.com/
[2] https://www.gerritcodereview.com/
[3] https://www.phacility.com/phabricator/

alternative order or empirical studies of code reviewers. By examining various review orders, this study seeks to get insights into actual code review practices. We specifically aim to answer the following research questions:

**RQ1**: How prevalent are out-of-order review comments and how strongly do they differ from the alphabetical order?
**RQ2**: To what extent can we observe meaningful orders in out-of-order reviews of pull requests?
  **RQ2.1**: How prevalent is the largest-diff first order?
  **RQ2.2**: How prevalent is the most-similar first order?
  **RQ2.3**: How prevalent is the test-first order?
**RQ3**: Does review coverage (as the proportion of reviewed files to all files in a PR) differ between alphabetical and non-alphabetical code reviews?
**RQ4**: Does the order followed impact the review outcome?

Prior research has shown that files appearing earlier in an alphabetical listing tend to receive more review comments. However, these findings are based on aggregate comment counts and do not consider the actual sequence in which reviewers leave comments. With RQ1, we aim to explore whether reviewers' commenting behavior aligns with alphabetical file order, by examining the chronological order and relative file position of comments. This allows us to revisit and validate previous findings from a more behavior-driven perspective and estimate how often reviews adhere to this order. RQ2 investigates whether reviewers follow alternative ordering strategies when reviewing files. We focus on three such strategies: largest-diff first and test first, both previously studied, and most-similar first, which we introduce based on anecdotal observations in the literature suggesting that reviewers may prioritize files most relevant to the PR's purpose. This question seeks to uncover the presence and frequency of more purposeful review flows beyond default file listings. To better understand the potential impact of review order, RQ3 considers review coverage—defined as the proportion of changed files that receive comments—as a proxy for thoroughness. By comparing coverage across alphabetical and non-alphabetical reviews, we assess whether order influences how comprehensively a change is examined. Finally, RQ4 explores whether review order has a measurable effect on the review outcome. Specifically, we look at the status of the first review round in accordance with whether or not the comments follow the alphabetical order.

We found that reviews that are not alphabetically ordered exhibit a stronger alignment with the alphabetical order as the number of commented files increases. This suggests that reviewers might use the default alphabetical order as a fallback strategy when faced with larger reviews. Additionally, non-alphabetically ordered reviews tend to comment on a significantly higher proportion of files compared to their alphabetically ordered counterparts. Our findings suggest a complex interplay between review ordering strategies, the extent of file coverage, and potentially, the rigor of the review process. Such insights may have important implications for optimizing code review practices and understanding the factors that influence review effectiveness. We propose a set of hypotheses to have a deeper investigation of our insights in future work.

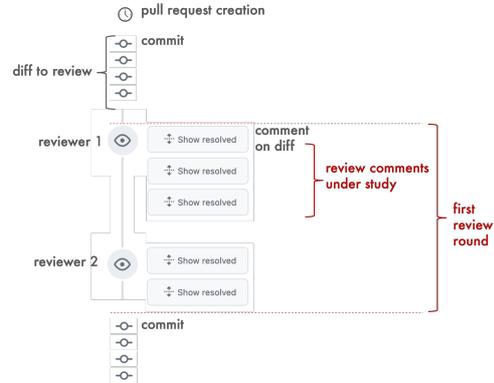

**Figure 1: Example of a first review round**

## 2 Study Design

We discuss the repository selection and explain how we collected candidate PRs, the corresponding review comments, as well as the studied orders and data needed to check them.

### 2.1 Repository Selection

As we aim to study code reviews on GitHub, we target open source projects having considerable contribution activity. Therefore, we target active projects from two popular programming languages, Java and Python, with at least 50 contributors and 1000 PRs. We used the seart GitHub Search [11] with the following criteria:

- Java and Python projects
- Recently active (last commit between 2023-01-01 and 2023-11-01, i.e., within the same year of our data collection)
- More than 1000 stars and a minimum of 50 contributors
- A minimum of 1000 PRs (showing that projects follow pull-based development model [18])

This search yielded a list of 315 Java and 497 Python repositories with a total of 3,406,262 PRs. From these, we randomly sampled 50 projects per programming language to obtain our set of 100 projects to analyze.

### 2.2 Candidate Pull Requests

For each repository in our sample, we first identified a list of PRs that are possibly relevant for our analysis. For this we queried the GitHub GraphQL API and considered each pull request that 1) has at least 2 code review comments, i.e., comments that refer to a certain line of code in the reviewed changes, and 2) has changes in at least 2 files of code. We cannot infer any ordering from PRs that have no comments, only a single comment, or only one file. In addition, we only considered PRs that were created after 2017-03-31 and closed or merged before 2023-11-01. For PRs created before 2017-03-31, the GitHub API returns different information encoding the code line that a comment is related to, which we could not reliably analyze. We chose the cut-off date of 2023-11-01 for the pull request completion to enable consistent re-runs of our analysis. The scripts used as well as our filtered dataset are available in our replication package[4]. For each candidate pull request in the retrieved list, we downloaded the pull request information, comments, and commits using the GitHub REST API. For this, we used a modified version of the ETRC tool developed by Heumüller et al. [15], which we made available[5].

Table 1: Code review orders derived from literature

| Order | Description | References |
|---|---|---|
| Test First | The files containing tests are shown before the files containing production code | [28, 30] |
| Largest-Diff First | The files with the largest count of added and removed lines come first | [3] |
| Usage Example | Code defining a natural entry point for the system comes first, i.e., the user interface, a command line interface or test cases | |
| Salient Class / Most Similar To Description | The central class targeted by the reviewed change is shown first. The central class is what is primarily modified, and causes the changes in other classes | [14, 16, 17] |
| Defect-Prone / Important | The changes most likely to contain defects, and/or those where the impact of defects would be worst are shown first | [6] |

### 2.3 Reconstructing Review Order

Code review may be conducted in multiple rounds during which one or more reviewers propose changes that are implemented by PR authors. Comment orders are sensitive to these review rounds and the number of reviewers involved. Thus, the entry point and order of a later review might be inspired by comments of the previous review round. Previous studies such as Spadini et al. [29] have shown that existing comments by other reviewers may influence the reviewer's behavior either positively by finding more instances of a bug that was pointed out, or negatively through availability bias. Hence, we only considered the first review round as depicted in Figure 1. This approach was also followed by Fregnan et al. [14] and Widyasari et al. [41] in their first commit analysis. We only considered comments made by users other than the pull request author, excluding bots, by filtering based on the user type field. We also excluded responses to previous comments and comments made by a second reviewer.

To detect out-of-orderness, i.e., review comment orders that do not adhere to the default alphabetical order, we ranked the creation time of the comments and the paths of the files targeted by a comment. Then, we calculated occurrences of *step backs*, where the rank of the creation time increases, but the rank of the file path decreases. This means the reviewer must have jumped back in the code diff to leave a comment. We considered each review with at least one of those step backs as out-of-order.

### 2.4 Meaningful Orders

Reviewers' choice of file order constitutes an integral part of their reviewing strategy as described by Baum et al. [6] in their interview study. Our choice of file orders to be investigated is informed by existing literature listed in Table 1 and the reported potential benefits of their adoption. In the following, we present the three meaningful orders we try to detect in more detail and explain our approach to identify each of them. We qualify these orders as meaningful on the one hand because they use file characteristics that the reviewer can choose to rely on in order to prioritize (i.e. whether a file is a test file, the size of the change, and the relevance of its contents to the overall change rationale) and on the other hand because they are simple enough and transparent to the reviewer following Baum et al.'s $6^{th}$ ordering principle, namely the use of rules that the reviewer can understand [6].

*Largest-Diff First.* Bagirov et al. [3] explored the impact of the file order on code review in the IntelliJ IDEA project. The authors compared the prioritization of problematic files under the default alphabetical order with the code diff order based on the number of changes per file. They found that ordering files based on their code diff outperformed the default alphabetical order when prioritizing problematic files. In this paper, we consider the code diff order described by Bagirov et al. [3] and denote it by largest-diff first. We consider the number of changed lines in each file by summing up the number of added and deleted lines. The commit compare endpoint of the GitHub REST API allowed users to query the number of added and deleted lines for each file. We used this end point to get the file changes made prior to the first review round.

*Most-Similar Files First.* Previous works such as the study conducted by Fregnan et al. [14] have suggested the use of alternative orders prioritizing the most critical files of a change. The authors refer to the work of Huang et al. [16] [17], who coined the concept of most *salient class*, described as the class causing the rest of the modifications in a commit. To identify such a class, Huang et al. [16] [17] developed a classifier using code coupling features, commit type features, code modifying features, and code semantic features. The latter consisted of a word embedding of the identifiers appearing in the class. We took inspiration from their work to develop a similarity-based order. This order takes into account code semantics and leverages transformer models trained on both source code and natural language to identify files whose code changes are most similar to the description of a PR. The underlying assumption of this order is that files with the highest semantic similarity with the goal of the pull request are the most relevant.

In order to approximate the relevance of the changes made to a certain file to the PR as a whole, we calculated the cosine similarity between the file changes on the one hand and the pull request title and description on the other. To achieve this, we concatenated the pull request title and description and calculate their embedding vector using the CodeT5+ [26] transformer model available on the HuggingFace platform [40]. CodeT5+ supports 9 programming languages including Python and Java and is trained on the stricter permissive subset of the github-code dataset. To obtain the embedding of the file changes, we considered each diff hunk

---
[4]https://github.com/abiUni/github_reviews_study/
[5]https://github.com/abiUni/mining_cr_data

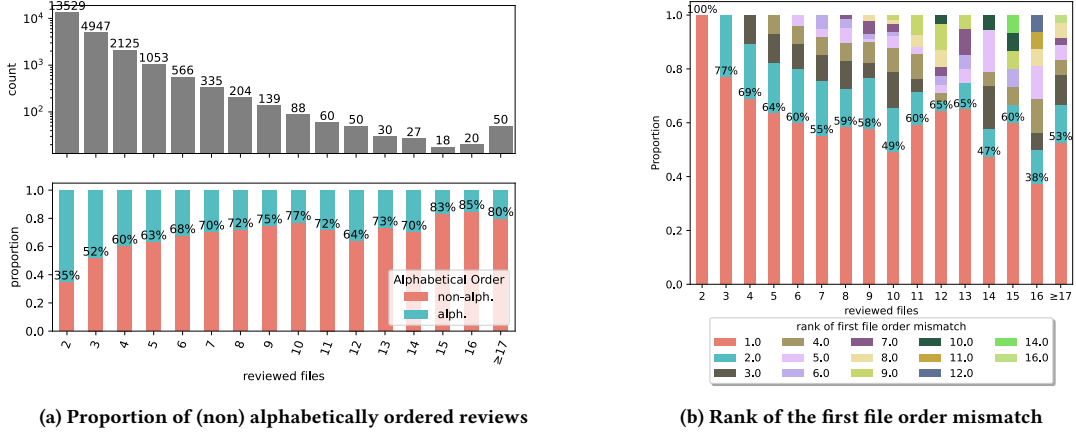

(a) Proportion of (non) alphabetically ordered reviews

(b) Rank of the first file order mismatch

Figure 2: Proportion of non-alphabetically ordered reviews and rank of the first file order mismatch

representing contiguous locations of line changes. For each diff hunk, we extracted and concatenated added and deleted lines and calculate their embedding using the CodeT5+ transformer. Lastly, to gain a single vector embedding per file, we aggregated all diff hunks corresponding to the file. We are then able to calculate the cosine similarity between the embedding vector of a file and the embedding vector of the pull request calculated previously using title and description.

*Test First.* In test-driven code review the reviewer starts inspecting changes in test files before transitioning to production files. Spadini et al. [28] conducted a within-subject experiment with 93 developers performing more than 150 code reviews with different strategies: test-driven code review, production-first code review, and only-production code review. Their results showed that developers practicing test-driven code review find the same proportion of defects in production code and proportionally more defects in test code. Consequently, TDR may contribute to the discovery of more defects in test code. Taking inspiration from their work, we consider the test-first order of files in our analysis. Following this order, files which have the keyword "test" in the stated path are ranked first and all other files are ranked second. Since we allow for rank ties, we do not distinguish in rank between the test files. The same holds for production files.

### 2.5 Kendell-$\tau$ Correlation

We aim to analyze the level of concordance between the selected meaningful orders and the observed out-of-order comments. Beyond analyzing strict matches between the observed file order of the comments and the reference orders, we use Kendall-$\tau$'s correlation coefficient with rank ties to assess the level of concordance between two ranks. Kendall-$\tau$ is a non-parametric test measuring the association of two variables based on the number of concordant and discordant pairs in paired observations. The test is used for instance to evaluate information retrieval systems by comparing a generated result ranking against a ground truth ranking [20].

For a set of observations $(x_1, y_1), ..., (x_n, y_n)$, any pair of observations $(x_i, y_i)$ and $(x_j, y_j)$ are said to be concordant for $i < j$, if the sort order of $(x_i, x_j)$ and $(y_i, y_j)$ matches, i.e., if $x_i > x_j$ and $y_i > y_j$ or $x_i < x_j$ and $y_i < y_j$ holds. In this case, we consider $X$ as the reference order and $Y$ as the observed reviewing order. Thus, the values $x_1, ..., x_n$ are the increasing rank of the files based on the given criteria. The Kendall-$\tau$ correlation coefficient varies in value between -1, corresponding to an inverted ranking between the two compared variables, and 1 indicating exact concordance. A value close to zero of the correlation coefficient indicates the absence of association between the compared rank variables.

The test requires two lists of ranks to be of the same length. To this end, we reconstructed the sequence of comments and the associated files. For each reference order, we calculated the rank of every unique file with at least one comment. We then assigned the rank of the corresponding file to each comment in creation-order. For example, for file A and observed rank 2 with 1 comment and file B with observed rank 1 and three comments (comments in order of occurrence), the observed sequence would be *2111*, while it would be *1222* in alphabetical order, given that B comes after A alphabetically. For the largest-diff first and most-similar first orders, we assigned ranks based on diff size and cosine similarity respectively, taking into account ties. For the test-first order, we only considered PRs containing both test and production files and assign rank 1 to all test files and 2 otherwise.

## 3 Study Results

### 3.1 Non-alphabetically Ordered Code Reviews

Our aim for RQ1 is to determine how many of the PR reviews deviate from the default alphabetical order. Out of a total of 23,241 PRs, in 10,377 (44.6 %) of the PRs the comments do not follow the alphabetical order. Figure 2a shows the proportion of non-alphabetically ordered reviews by number of reviewed files in a PR. Note that PRs with 17 or more files are grouped into the "≥17" category due to their low counts. We can observe an increasing trend of non-alphabetically ordered reviews as the number of reviewed files rises.

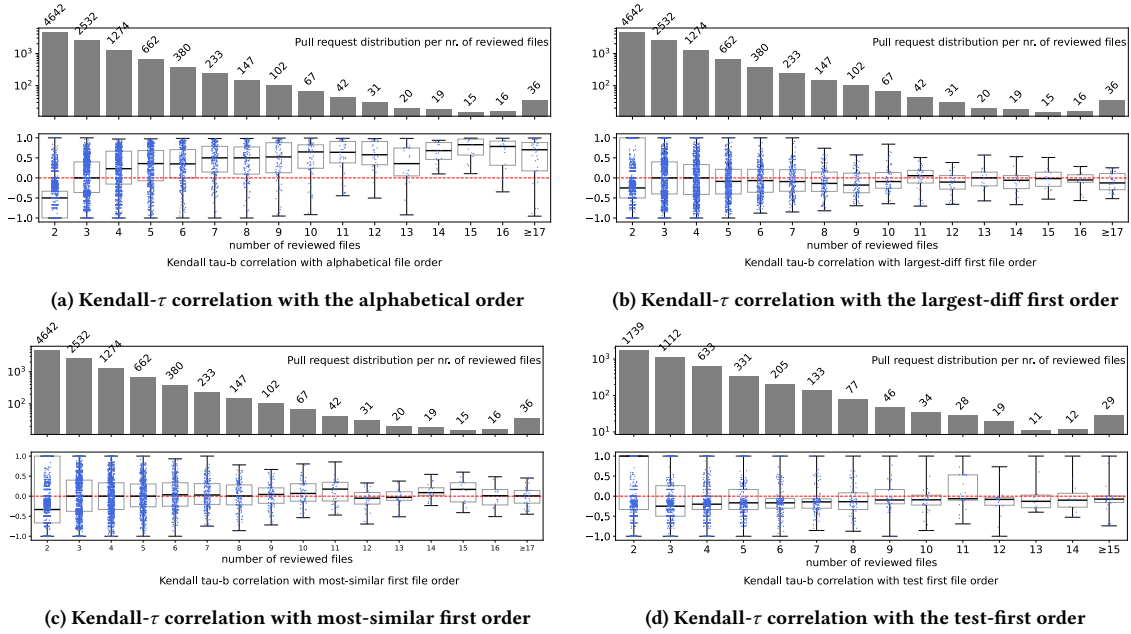

Figure 3: Kendall-$\tau$ correlation with file orders

Starting from 3 reviewed files, non-alphabetical reviews make up the majority with more than 50% of all PRs.

We then take a closer look specifically at the set of non-alphabetically ordered PRs. To investigate their alignment with the alphabetical order, we calculate the Kendall-$\tau$ correlates depicted in Figure 3a. The figure shows an increase in overall alignment with the alphabetical order for a growing number of reviewed files, with a consistently positive correlation for four or more files. This suggests an overall adherence to alphabetical order with minor deviations, further supported by Figure 2b, which highlights the rank of the first mismatch between the alphabetical and the observed order. Figure 2b shows that deviations from the alphabetical order tend to occur later in the sequence as the number of reviewed files increases. For example, for 3 reviewed files, 77% of reviewers begin by commenting on a file other than the first in the alphabetical order. This percentage drops to 49% for 10 reviewed files.

> **Answer to RQ1: How prevalent are out-of-order review comments and how strongly do they differ from the alphabetical order?** We observe that software engineers do not follow alphabetical order in their reviewing of nearly 45% of the PRs. When code reviews are segmented by the number of files receiving comments, deviations from alphabetical order increase with the number of files reviewed. However, as the number of reviewed files grows, these non-alphabetical comments also show a stronger correlation with the alphabetical order, indicating a general adherence with minor deviations.

### 3.2 Prevalence of Meaningful Orders

*3.2.1 Largest-Diff First.* We examine whether out-of-alphabetical-order reviews follow a largest-diff-first approach, where files with the most changes are commented on first. Among the 10,377 non-alphabetical PR reviews, 2,134 (20.6%) strictly follow the largest-diff-first order. Using a more flexible alignment criterion (Kendall-$\tau$ correlation $\geq 0.5$), we find that 2,584 (24.9%) of these reviews align with this order. Figure 4a shows that the majority of reviews following the largest-diff-first order occur when *few files* are commented on. As the number of commented files in a PR increases (note the logarithmic scale), the proportion of reviews following even a soft largest-diff-first order steeply declines.

To investigate how closely the review orders align with a largest-diff-first order, we present their correlations in Figure 3b. For code reviews with a small number of files, the correlation values are widely distributed—from perfect alignment with largest-diff-first (*corr*=1), through no correlation, to a reverse largest-diff-last order (*corr*=−1). As the number of commented files increases, these correlations converge toward the center, indicating a shift to no correlation with this order.

*3.2.2 Most-Similar Files First.* We find that 1,827 (17.6%) of non-alphabetical PR reviews adhere to a strict most-similar-first order. Figure 4b shows the number of PRs that follow both strict and soft most-similar-first orders by the number of reviewed files. Under the soft definition (*corr* $\geq 0.5$), 23.5% (2,442) of non-alphabetical reviews exhibit strong alignment with this order. However, as with largest-diff-first, adherence to this order decreases to zero as the number of reviewed files increases, dropping off at 7 files for the

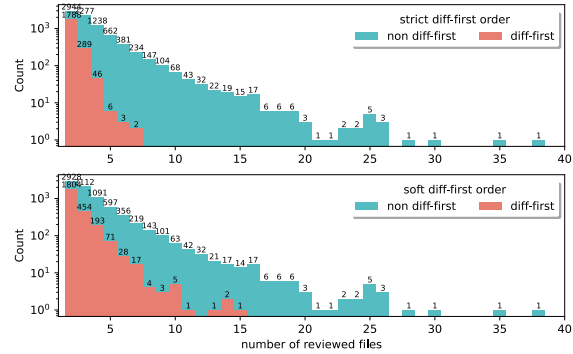

(a) (Non) diff size-based ordered code reviews

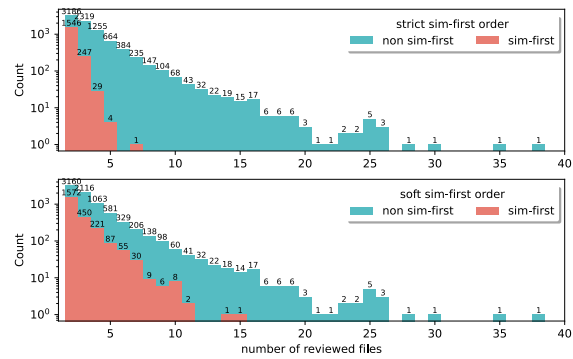

(b) (Non) similarity-based ordered code reviews

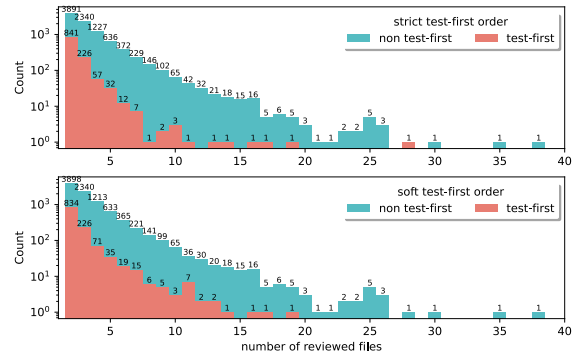

(c) (Non) test-first ordered code reviews

Figure 4: Distribution of reviews per file order, soft orders correspond to corr. of ≥0.5, while strict corr. = 1

strict definition and 10 files for the soft one. A closer look at Kendall-$\tau$ correlation scores in Figure 3c reveals that average correlations only slightly deviate from zero towards positive values for certain counts of reviewed files (e.g. 11 reviewed files).

*3.2.3 Test First.* The test-first order prioritizes the review of test files over production files and is only relevant for PRs that contain both. Consequently, we narrow our sample to those PRs, resulting in a total of 4,059 PRs. In contrast, 5,902 PRs contain only non-test files, while 416 contain only test files. Within this sample, 1,188 PR reviews adhere to the strict test-first order, representing 29% of the non-alphabetical PRs that include both test and non-test files. Looking at the Kendall-$\tau$ correlation of non-alphabetical PRs in the above defined sub-sample and their alignment with the test-first order, we can see that the coefficient is consistently negative with varying number of reviewed files (from 3 to 15 and up). This negative correlation might be indicative of an overall preference for reviewing production files first. In fact, when taking a closer look at the PRs not following a strict test-first order, we see that the majority start with production files as depicted in Figure 5. We should note that our definition of test-first order entails considering all reviewed test files before considering production files, thus merely starting the review from a test file does not meet the requirements for this order. In a related context, Beller et al. [8, 9] conducted a large scale field study monitoring developers' actions in the IDE to investigate their testing behavior. The authors observed that no developer follows Test-Driven Development (TDD) continuously. They advocate for a Test-Guided Development instead, i.e. loosely guiding the development with the help of tests. Similarly, our findings suggest that reviewers might use tests to guide the review. This assumption, however, is subject to scrutiny and a more detailed investigation in future work.

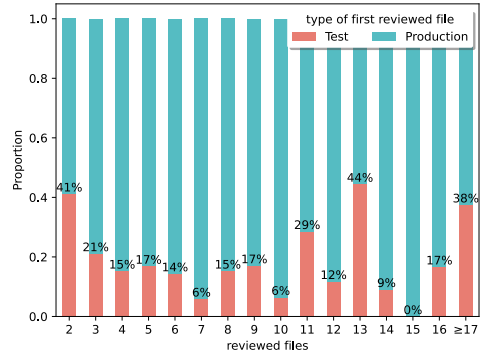

Figure 5: Type of file at the start of the review (entry point)

> **Answer to RQ2: To what extent can we observe meaningful orders in out-of order reviews of pull requests?** None of the studied meaningful single-concern orders can fully explain the reviewers' deviation from the alphabetical file order. Nevertheless, we can identify non-alphabetical reviews that follow (soft variants of) each meaningful order, indicating that some reviews do, in fact, follow these orders. However, for large numbers of files discussed during code reviews, the correlation with meaningful orders clearly trends towards zero.

### 3.3 Review Coverage and Reviewing Order

We calculate the proportion of reviewed files to all files in a PR as an indicator of review coverage. Generally, the number of comments and reviewed files has been taken as an indictor for review quality in experimental studies with reviewers [24]. Since the GitHub UI does not allow a user to show more than 300 files, we excluded 98 PRs with 300 or more files. Figure 7 shows the distribution of the review effort for PRs following the alphabetical order in comparison to non-alphabetically ordered PRs. The average proportion of reviewed files is higher (42.1%) for non-alphabetically ordered PRs compared to alphabetically ordered ones (39.0%). Applying a non-parametric Mann-Whitney U test, we found that the differences of averages is significant ($p<0.001$).

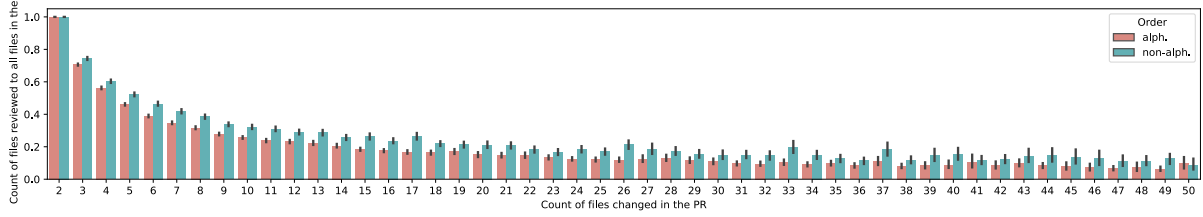

Figure 6: Distribution of the proportion of reviewed files to all files submitted for review in a pull request

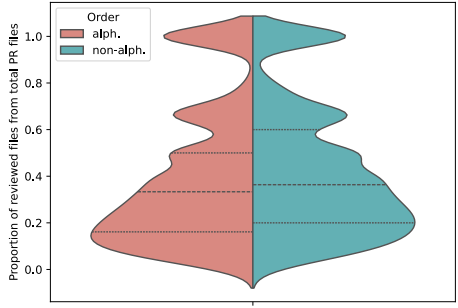

Figure 7: Proportion of reviewed files to total files

Figure 6 shows a detailed view of the average proportion of reviewed files for PRs containing changes of up to 50 files. We can see that the proportion of reviewed files is 100% for PRs with two files and drops with increasing number of files to be reviewed. Furthermore, for nearly all file counts, the average proportion of reviewed files is consistently higher when reviews are conducted in non-alphabetical order.

> **Answer to RQ3: Does review coverage (as the proportion of reviewed files to all files in a PR) differ between alphabetical and non-alphabetical code reviews?** As the number of files changed in a pull request grows, the proportion of commented files decreases. For PRs with more than 2 files, we observe that, when more files are reviewed the review is more often in non-alphabetical order. Developers comment on a significantly higher proportion of files in non-alphabetically ordered code reviews.

### 3.4 Review Outcome and Reviewing Order

Each review round on GitHub can be assigned a status by the reviewer as shown in Table 2. Since we only consider closed PRs, the status "PENDING" is not present in our sample. We were able to retrieve the review status of 23,145 PRs using the GitHub GraphQL API, with 96 PRs missing a valid status. Figure 8 (a) shows the percentage of non-alphabetically ordered PRs by review status. Non-alphabetical PRs make up 47% of commented reviews, 44% of reviews requesting changes and 43% of all reviews that were dismissed. These percentages are comparable to their overall representation in the sample (44.6%). However, they make up a slightly smaller percentage (36%) of approved reviews overall.

Figures 8(b)–(d) detail the percentages per status for each of the investigated orders compared to the overall non-alphabetically ordered PRs. We should note that the sample is reduced for the test-first order as it includes PRs containing both test and production file changes. The representation of meaningful orders with the status APPROVED among the total number of non-alphabetically ordered PRs is higher than their representation of all other statuses (19% for most-similar first, 24% for largest diff-first and 34% for test first).

> **Answer to RQ4: Does the order followed impact the review outcome?** Among approved reviews, the share of non-alphabetically reviewed PRs (36%) is lower than that of alphabetically ordered PRs (64%).

## 4 Discussion
### 4.1 Interpretation of the Results

Through our analysis of review order, we have found that a considerable amount, nearly half, of PRs do not follow the default alphabetical order presented in the GitHub UI. On average, an increase in correlation with the alphabetical order as the number of files per pull request increases is detected. By analyzing the file rank of the first mismatch, we found that for larger reviews, the mismatch occurred later in the review compared to smaller reviews. These results suggest that reviewers potentially start reviewing files following the order shown by the tool and deviate from the default order later on. We can therefore hypothesize that the default order is used as a fallback strategy for larger reviews.

Table 2: Distribution of review outcomes in the PR sample

| Status | Count | % | Description |
|---|---|---|---|
| COMMENTED | 13,625 | 58.8 | Informational review |
| CHANGES REQUESTED | 6,293 | 27.2 | Requested author changes |
| APPROVED | 3,020 | 13.0 | Changes approved, merge allowed |
| DISMISSED | 207 | 0.9 | Change request dismissed |
| **Total** | **23,145** | **100** | |

Moreover, while the investigated meaningful orders can explain some of the deviations from the alphabetical order, they also become less applicable as the number of reviewed files increases. This implies that developers likely use a combination of strategies or develop their own heuristics to navigate larger reviews. Hence, reviewers need improved guidance and file overview tools to help them find suitable entry points within complex reviews based on pieces of information relevant to their preferred strategy. Based on their interview study with reviewers, Baum et al. introduced the concept of "The Macro Structure", which emphasizes that reviewers initially need an overview of the changes before determining an entry point. They reported various tactics that reviewers employ, such as beginning with simpler changes, starting from natural entry points like graphical interface components, or prioritizing the most significant changes or unfamiliar areas [6]. Notably, current tooling such as GitHub offers limited support in providing such overviews, which reviewers rely on to guide their starting points. Our findings partly align with Baum et al.'s documented review tactics. Their observation that some reviewers prefer to begin with simpler changes helps explain the slightly negative average correlation with the largest-diff-first ordering (Figure 3b). Furthermore, GitHub's interface provides limited context about the relative importance of file changes within PRs. Consequently, following a similarity-based review order requires reviewers to have prior knowledge of the codebase.

Our review coverage analysis has shown that the proportion of reviewed files to all files in a pull request is significantly higher in non-alphabetically ordered reviews. This leads us to the hypothesis that deviating from the default alphabetical order coincides with higher code review coverage and should therefore be encouraged and supported by existing tools. On a similar note, reviews following a non-alphabetical order resulted in slightly fewer approvals in the first review round, suggesting that those reviewers were more critical of the submitted change. Further analysis of the review comment content is needed to confirm or refute this interpretation.

### 4.2 Recommendations

Imposing an arbitrary order, such as the alphabetical order, during code reviews may reduce code review effectiveness. Our findings show that no single ordering approach dominates, and therefore, we advocate for improving the availability and presentation of key information to assist reviewers in developing an effective review strategy. Previous works, such as those by Baum et al. [6], have established principles for suggesting a file order and emphasized the importance of transparently communicating the rationale behind a recommended order. This is reflected in the sixth principle of their work, which states that failing to explain the reasoning behind the order can *"break his [the reviewer's] line of thought"* and lead to disorientation. Earlier works such as Maalej [23] advocate for the integration of context-based information within development tools to alleviate the burden of searching for scattered information.

To better support meaningful review strategies, the user interface of code review tools could be enhanced with visual indicators, such as displaying diff sizes in the file tree thus providing a better overview of this indicator rather than its current placement in the diff view requiring scrolling between files. Additionally, file-level summaries of changes would help reviewers understand the overall structure of the change and prioritize their review accordingly. Rather than just recommending a specific order, offering hints about potential entry points could be beneficial, particularly for larger reviews, as reviewers often start from the first file in alphabetical order. Finally, recognizing that developers may need to use a combination of strategies instead of relying on a single order, review tools should allow reviewers to freely reorder the file list with ease.

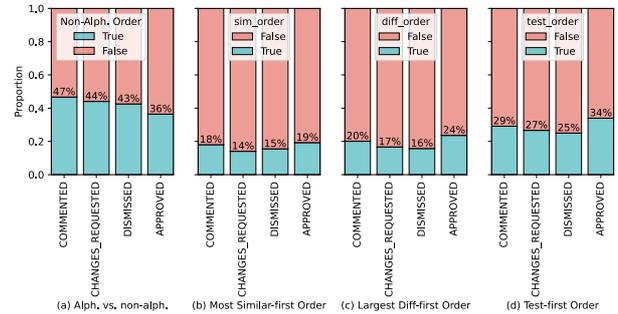

Figure 8: Distribution of review status by observed order

### 4.3 Hypotheses for Future Work

Based on the findings of this study, we propose the following hypotheses for future exploration:

**H1. Alternative orders have an impact on review efficiency** Providing reviewers with alternative navigation orders, such as largest-diff first or test first, may improve review efficiency and coverage, particularly for larger PRs.

**H2. Review order correlates with comment depth** Reviews conducted in a non-alphabetical order are hypothesized to yield deeper, more critical feedback due to increased attention on files deemed significant by the reviewer.

**H3. Reviewer familiarity with the code correlates with order preference** Reviewer familiarity with the codebase may influence their preference for certain navigation strategies, such as similarity-based ordering.

**H4. Enhanced tool support affects the choice of review strategies** Augmenting code review tools with features like file-level summaries and diff-size visualizations is hypothesized to reduce reliance on default alphabetical order and encourage more meaningful review sequences.

**H5. Reviewers use a combination of strategies in large reviews** For PRs with a large number of changed files, reviewers are hypothesized to combine multiple strategies (e.g., largest-diff first and similarity-based) rather than adhering strictly to one.

## 5 Related Work

### 5.1 Code Comprehension and Code Review

To deliver a code review verdict, reviewers first need to understand the change in its context and assess its necessity and quality. Thus, code comprehension is fundamental to every code review activity.

Known strategies such as bottom-up, top-down, and integrated comprehension strategies may be used by reviewers as they build their understanding of the change [31][39][10][19]. Early studies on software inspections have proposed reading techniques including "use-based reading" by Thelin et al. [33] and "functionality-based reading" by Abdelnabi et al. [1], which use a top-down approach to trace use cases and functionalities. Dunmore et al. [12] proposed "abstraction-based reading" using a bottom-up view of a program through summaries. However, these early findings are based on reading one code version instead of reading code changes, which is the most common representation to-date.

In fact, change comprehension is a specialized case of code comprehension primarily due to the diff representation. In their study of information needs during code review, Pascarella et al. [25] found that reviewers process diverse information about the code change, its rationale, its place within the larger code base, and its impact on software quality. Bacchelli and Bird report in their interview study with Microsoft developers that no other code review challenge has emerged as clearly as change understanding [2]. Tao et al. [32] studied how software engineers understand code changes specifically. Their results show that rationale is the most important information reviewers need to understand the change, and they primarily rely on the change description to meet this need. Additionally, reviewers tend to rely more on their own knowledge of the code rather than on historical data, such as defect proneness. When asked about desirable tool support features, respondents mentioned IDE-like features to help navigate the diff.

Bacchelli and Bird coined the term "A Priori Understanding" to describe the observation that some participants took the time to read the textual description carefully, while others went directly to a particular changed file [2]. These participants explained that it was easier for them to understand a change if they are owners or very familiar with the files being changed. The authors further confirmed this finding in a survey, in which the respondents stated taking longer to review files they are unfamiliar with.

### 5.2 File Ordering in Code Review

Code review performance has commonly been measured in terms of the number of defects found. Many factors have been theorized to influence performance including the attention the reviewer pays each changed file. One such factor recently gaining attention is the organization of files during review. The work of Baum et al. is seminal to investigating the relation between the alphabetical file ordering and commenting activity during code review [6]. The authors used a theory-generated methodology by combining results from multiple sources, namely: analysis of log data from 292 review session in industry, task-guided interviews with professional developers, related work from cognitive science and program comprehension, and a task-based survey conducted with 201 reviewers. The main result of their study is a set of principles describing order properties to help improve review efficiency [6].

A more recent study by Fregnan et al. [14] tests the hypothesis stating that the relative position in which a file is presented biases the code review outcome. The study refers to psychological factors such as attention decrement and working memory as plausible explanations for the potential impact a file's position may have on the code review outcome. To test their hypothesis, the authors triangulate complementary evidence based on an analysis of PRs from 138 Java projects on GitHub and an online experiment measuring participants' performance in detecting two unrelated defects in separate files. The authors found that files shown earlier receive more comments compared to files shown later even when controlling for possible confounding factors such as the number of added lines in a file. They consider the number of comments as an indication of the attention the file receives during the code review.

A recent work by Bagirov et al. [3] studied the impact of the alphabetical order of files in the IntelliJ IDEA project. The authors compare the performance of alphabetical ordering against a *Code Diff*, which prioritizes files with the largest number of changes. They first identify the most problematic files defined as the ones requiring more attention from the reviewers. They subsequently use Reciprocal Rank (MRR) to represent the positions of the first problematic files, $precision_k$ to calculate the number of problematic files among the top k files, and Discounted Cumulative Gain (DCG) to get the overall ranking quality. Their results show that the Code Diff order significantly outperforms an alphabetical ordering with regard to MRR and DCG. Both Fregnan et al. [14] and Bagirov et al. [3] demonstrate that files positioned earlier in the alphabetical order tend to receive more comments. Our study differs by taking a closer look at the actual commenting sequence followed by the reviewers, rather than relying on a summative approach that relies solely on the number of comments. Moreover, we attempt to find actual evidence of alternative orders that the reviewers might have incorporated in their reviewing strategy, taking inspiration from the largest-diff first order studied by Bagirov et al. [3].

We also draw from Olewicki et al. [24] in our used of diff embeddings for the most-similar first order. The authors leveraged text embeddings using Bag-of-Words and LLMs (Bloom from Big Science) on added and deleted lines as well as review process features to predict files needing comments, revisions, or attention as hot-spots (need either comments or revisions). During an evaluation with 29 expert reviewers, they found that re-ordering the files based on their approach resulted in 23% more comments, and an increase in participants' file-level hot-spot precision by 13% and recall by 8% compared to the alphanumeric ordering. However, the authors observed that the order chosen by the reviewers varies a lot based on their experience with the files and reviewing habits.

### 5.3 Code Review Automation

Recently, an increased focus has been put on automating the code review process. Early work by Tufano M et al. [35] proposed using neural machine translation (NMT) to generate comments like a human reviewer would. Later, Tufano R et al. [38] suggested the use of NTM for two tasks: a *code-to-code* task generating the modified code, and a *code and comment-to-code* task to translate a review comment into code modifications [36]. Subsequent work by Thongtanunam et al. improved on their original approach by proposing AutoTransform. AutoTransform uses of Byte-Pair Encoding (BPE) to deal with new tokens [34]. The popularization and increased availability of transformer-based models through the HuggingFace platform [13], saw increased use of Text-To-Text Transfer Transformer models, specifically the T5 model family, to solve software engineering problems. Tufano R et al. [37] trained a T5-based model architecture to automate the *code-to-code* task and the *code-to-comment* task. This architecture was also used by Li et al. [21] for the CodeReviewer model, capable of generating a code diff quality estimation in addition to review generation and code refinement. Recently, The work of Lu et al. [22] proposed LLaMA-Reviewer, a parameter-efficient fine-tuning framework using less than 1% of the trainable parameters and achieving better results than prior work. Although code review automation plays a key role in reducing the workload for humans, reviewers ultimately remain responsible for deciding whether to merge a change into the codebase. Understanding code changes therefore continues to be a challenge, and with it the optimal representation of files.

## 6 Threats to Validity

**Internal validity.** We sampled code reviews with at least two commented files to assess the review order, excluding PRs with two or more changed files but fewer than two comments. This sample may therefore not represent low-effort PRs. Additionally, we only considered the first review round by a single reviewer, not accounting for multi-round review processes. As such, our sample may not represent the review process in practice. However, this simplification is necessary to comprehend the reviewing strategy with respect to the order of the files in the absence of external influence, e.g., from previous comments.

While analyzing meaningful orders, we did not calculate the accidental alignment between orders. By definition, the orders are of different granularity levels, e.g., the test-first order confers the same rank to all test files, while the most-similar first order assigns ranks based on the cosine similarity coefficient of each file to the pull request title and description. We also do not study the accuracy of the embeddings for the most-similar first order. However, we employed a state-of-the art model used in previous research in the context of code review comment generation. Even as overlaps might inevitably occur, the orders we chose are conceptually distinct.

Furthermore, the code reviews studied might be affected by several confounding factors that we do not treat in our analysis. Among these factors, the familiarity and experience of the reviewers with the code base and the review guidelines in place can play a role, as is the case for code comprehension at large [27]. Also we do not treat the content of the comments themselves. Previous works such as Beller et al. [7] have shown that code review comments focus more on maintainability rather than functional defects with a 75:25 ratio. Functional defects may require more attention from the reviewer and could exhibit different commenting intensity from maintainability defects.

**Construct validity.** This study relies on the order of comments as a proxy for the order in which reviewers navigated the files during code review. While this approach enables large-scale analysis, it introduces potential inaccuracies since reviewers might examine files without leaving comments. We consider the presence of a comment as indicator of the order in which the reviewer read the changed files. However, without any other evidence of the actual reading order, it remains possible that some reviewers perform an initial pass through the code before leaving their comments. While we cannot rule out this scenario completely, we recognize that it requires additional effort by developers and is therefore likely negligible for PRs demanding moderate to high effort.

We considered metrics such as the Needleman-Wunsch algorithm used to calculate alignment of protein or nucleotide sequences. However, the sequences in our case are different as we compare a reference order of fixed length against an observed order of arbitrary length. We therefore opted for Kendall-$\tau$'s correlation coefficient to assess the alignment with reference orders.

**External validity.** We chose our sample among popular Python and Java repositories on GitHub following the best practice in software engineering research. Among these repositories, we chose a random sample of 50 repositories for each programming language. Thus, our findings may not generalize well to other languages or application domains with different review practices or project cultures. Further, our study is closely tied to the GitHub workflow, which might not generalize to other reviewing tools, e.g., Gerrit.

## 7 Conclusion

In this study, we analyze the first review round from 23,241 PRs stemming from 100 GitHub projects. We investigated to what extent the reviewers followed the default alphabetical order based on the positions of the comments they authored. Our results reveal that nearly half of the reviews deviate from the default alphabetical order. However, non-alphabetical reviews correlate more strongly with the alphabetical order as the number of commented files increases, suggesting that this order might be used as a fallback strategy with increasing complexity. Non-alphabetical reviews also cover a significantly higher proportion of changed files. While we identified instances of meaningful orders, none of them fully explain the observed patterns, particularly in larger reviews. These findings underscore the need for enhanced tool support in code review interfaces to accommodate diverse review strategies and possibly improve review effectiveness. In the future, we would like to study more closely the effect of reviewers' experience with the code base, their reviewing habits, and the project guidelines on the chosen review order. These effects are more appropriately investigated in the context of a study with human subjects to externalize the rationale behind the choice of order.

## 8 Acknowledgments

This work was partially supported by the Zonta International Women in Technology (WIT) Scholarship and by the Dutch science foundation NWO's Vici grant "TestShift" (VI.C.182.032). Additional sponsoring came from the Swiss National Science Foundation through SNSF Grant 200021M 205146 as well as from the Deutsche Forschungsgemeinschaft (DFG, German Research Foundation) - Project Number: 166725071.